# Alloy Design for Mechanical Properties: Conquering the Length Scales

Irene J. Beyerlein[a,*], Shuozhi Xu[b], Javier LLorca[c,d], Jaafar A. El-Awady[e], Jaber R. Mianroodi[f,g], Bob Svendsen[f,g]

[a]Department of Mechanical Engineering, Materials Department, University of California, Santa Barbara, Santa Barbara, CA 93106, USA

[b]California NanoSystems Institute, University of California, Santa Barbara, Santa Barbara, CA 93106-6105, USA

[c]IMDEA Materials Institute, C/Eric Kandel 2, Getafe, Madrid 28906, Spain

[d]Department of Materials Science, Polytechnic University of Madrid. E. T. S. de Ingenieros de Caminos., Madrid 28040, Spain

[e]Department of Mechanical Engineering, Whiting School of Engineering, The Johns Hopkins University, Baltimore, MD 21218-2682, USA

[f]Microstructure Physics and Alloy Design, Max-Planck-Institut für Eisenforschung GmbH, Düsseldorf, D-40237, Germany

[g]Material Mechanics, RWTH Aachen, Aachen, D-54062, Germany

Predicting the structural response of advanced multiphase alloys and understanding the underlying microscopic mechanisms that are responsible for it are two critically important roles modeling plays in alloy development. An alloy's demonstration of superior properties, such as high strength, creep resistance, high ductility, and fracture toughness, is not sufficient to secure its use in widespread application. Still, a good model is needed, to take measurable alloy properties, such as microstructure and chemical composition, and forecast how the alloy will perform in specified mechanical deformation conditions, including temperature, time, and rate. In this bulletin, we highlight recent achievements by multiscale modeling in elucidating the coupled effects of alloying, microstructure, and the dynamics of mechanisms on the mechanical properties of polycrystalline alloys. Much of the understanding gained by these efforts relied on integration of computational tools that varied over many length and time scales, from first principles density functional theory, atomistic simulation methods, dislocation and defect theory, micromechanics, phase field modeling, single crystal plasticity, and polycrystalline plasticity.

**Introduction**

For at least a century, modeling has played an integral and critical role in alloy design. 2D maps, equations or rules are some of the most popular and utilized forms of modeling. Ashby maps have facilitated materials selection, enabling comparisons between classes of alloys against measurable performance indices, such as specific stiffness and yield strength.[1] The Hall-Petch scaling law has been used through the years to relate material strength/hardness to the size of the grains in a polycrystal.[2,3] The Hume-Rothery rules are widely used to guide the choice of alloying elements to indicate whether the alloy will be single or multiphase.[4]

While the attractiveness and insight afforded by straightforward rules, expressions, and maps are greatly appreciated, the needs and challenges to develop better material models still persist. Alloy design has stretched to encompass more than simply choosing which and how many solute atom(s) to add. Additions of just a few percent of alloying elements can profoundly affect the activation of nanoscale mechanisms (operating over Å – nm scales) when the alloy is deformed. With heat treatment and mechanical processing, these atomic-scale element additions can affect the evolution of the internal microstructure (characteristic length scales > nm). The resulting alloy microstructures are often complex, consisting of multiple phases with each phase containing more than one type of precipitates, particles, and interfaces. Changes in any of these features can, in turn, influence a broad suite of critical bulk structural properties (samples > mm), such as strength, ductility, fatigue, fracture, and creep. In order to continue to benefit the design and optimization of alloys, material models need to be multiscale, spanning from atoms to the continuum.

A full-spectrum, "atoms-to-continuum", multiscale material models (MMMs) for alloy design, however, does not currently exist. Atomic-scale models exist that treat a polycrystal as a collective arrangement of atoms. Continuum models also exist that treat the sample or structure as a deformable continuum. The modeling components that are missing lie in the intermediate scales, between nm and mm. This set of scales is collectively referred to as the *mesoscale*. Crossing the vast mesoscale gap is where MMM meets its greatest challenges.

**Figure 1** displays a current view on mesoscale MMM for advanced alloys. At its core (central circle), mesoscale MMM intends to benefit alloy design by aiding in creating, understanding, and designing new materials. These MMM efforts tend to pursue one of three goals (intermediate ring). Some MMMs are built to improve and discover deformation mechanisms. Other MMMs are developed to simulate microstructural evolution under a given manufacturing process so they can identify pathways to achieve target microstructures. Last are MMMs that focus on determining microstructure-material response relationships in efforts to address the question, "which microstructures are the 'right' ones?". At the periphery of the MMM (**Figure 1**) lie the multitude of mesoscale multiscale models to date that treat more than one length scale in the mesoscale spectrum. This short communication aims to highlight some of these MMM achievements for alloy design.

**Coupling the thermodynamics and mechanics of precipitation**

Precipitation hardening is well-established as one of the most efficient strategies to increase the yield strength of alloys. Precipitates are metastable intermetallic particles with sizes ranging from a few to a few hundred nm that appear during aging. Improvements in alloy strength depends sensitively on the precipitate size, shape, and spatial distribution, as well as how moving dislocations interact with them. In their article in this issue, Nie and Wang[5] highlight many recent achievements in precipitation design for improved lightweight alloys.

Multiscale modelling strategies can aid in the design of novel precipitation hardened alloys by replacing costly experimental trial and error approaches. Recently, one such MMM was built for the precipitation process in Al-Cu alloys.[6-11] At its foundation is an analysis of thermal stability for each kind of precipitate, which is accomplished by first principles calculations of the Helmholtz free energy.[7] Precipitate nucleation and growth are then predicted, respectively, via classical nucleation theory and phase-field mesoscale simulations (**Figure 2**).[7,8] The former is used to predict the initial dimensions and shape of the precipitate nucleus. When nucleation is homogeneous, it calculates the chemical, interfacial and elastic driving forces using computational thermodynamic data, first principles calculations, and the lattice correspondence during the transformation. In the likely situation of heterogeneous nucleation—that is, when other precipitates, such as dislocations or other defects, can alter the nucleation process—defect stresses can be taken into account in the multiscale framework. Once the initial dimensions and shape of the precipitate nucleus have been determined, precipitate growth to its stable shape is calculated using the phase-field method.[8]

Multiscale simulations from this strategy can reveal the stability and growth sequences of precipitation, as well as the expected final precipitate microstructure. **Figure 2**, for instance, demonstrates the profound effect dislocations can have on the final equilibrium shape of the precipitates, being flat disks when nucleation is homogeneous vs. cones and platelets with morphologically rough edges when it is heterogeneous.[7,8]

**Chemo-mechanical interactions between gliding dislocations and precipitates**

The amount of strengthening provided by precipitates to an alloy depends heavily on several atomistic and mesoscale aspects of the mechanisms used by dislocations to bypass these precipitates. While a few bypass processes have been proposed and studied, the mechanism is usually not known *a priori* since it depends on a strong coupling among the geometric, chemical, and mechanical properties (e.g., crystal structure, composition, stiffness, lattice orientation) of the precipitates.

Ni-Al-based superalloys provide one prominent example of the need to consider both the mechanical and chemical aspects in processes of dislocation/precipitate interactions. While alloying provides for a microstructure consisting of precipitates, e.g., the $\gamma'$ phase, further improvements in strength could be gained by adding solute elements, such as W, Re, Co, and Cr. However, solute segregation to defects, such as dislocations and stacking faults[12-15] could instead lead to precipitate dissolution, enhanced directional coarsening ("rafting"), and degradation of mechanical properties. Due to the chemo-mechanical couplings involved, calculation of the mechanisms and associated critical energies and stresses required by dislocations to bypass the precipitate call for a multiscale strategy.

In the last few decades, multiscale models combining atomistic calculations and phase field (PF) modeling have been developed to simulate dislocation motion.[16-18] To treat alloys, the energetics used in the PF-dislocation simulations have been advanced to include chemo-mechanical couplings.[19,20] In these models, the role morphology and chemistry play in the mechanisms that underlie dislocation/interface interactions are encompassed in energetic terms used in the master energy function of the PF method.

**Figure 3** shows a result from one recent example in this class.[20] The MMM combines a general phase-field-based chemo-mechanical methodology[21] with PF-dislocation modeling to simulate a dislocation shearing a precipitate in Ni-Al-Co. Under an applied stress, the dislocation glides in the matrix toward the precipitate. Before reaching the precipitate, the dislocation has dissociated into two Shockley partial dislocations and is decorated by Co (**Figure 3a**). At the $\gamma$-$\gamma'$ interface, these partials recombine into a perfect edge dislocation before they enter and shear the precipitate. As the dislocation glides into the $\gamma'$ precipitate, it deposits Co on the $\gamma$-$\gamma'$ interface (**Figure 3b**) as well as drags Co into the precipitate (**Figure 3c**). Ultimately, as seen in **Figure 3d**, Co segregation enriches the faults left by the shearing process and depletes Co in the precipitate $\gamma'$ matrix.

The coupling between dislocation shearing and solute segregation could promote precipitate dissolution, directional coarsening ("rafting"), loss of strength, and lifetime reduction. The insight given by these multiscale simulations can be useful towards optimizing alloying composition to delay or inhibit such effects.

**Moving dislocations across scales**

A computational approach that can provide understanding and prediction of the deformation of alloys is three-dimensional discrete dislocation dynamics (3D DDD) simulations. This mesoscale technique was developed to model plastic deformation as the result of the collective motion of many dislocations gliding in a crystal.[22-25]

In recent years, DDD has been applied to model microstructure/response relationships in superalloys and in particular, the effect of several dislocation/precipitate interactions in a crystal.[26] **Figure 4a** shows the predicted response of Ni-base superalloys as a function of the size of the cuboidal precipitates within the single crystals of the polycrystalline alloy. In these calculations, this mesoscale technique advantageously accounts for randomness in the distribution and shapes of the precipitates in the crystal, the evolution of a collection of dislocations, and the numerous precipitate/dislocation interactions. The model reveals that the effect is due to a change in the dislocation/precipitate bypass mechanism with increasing precipitate size (see **Figure 4(c-d))**.

3D DDD simulations have successfully quantified many other important microstructure/strength relationships, such as a nearly independent relationship with grain size (compared to the strong precipitate size effect), linear increase with precipitate volume fraction, and an approximate square-root relationship with anti-phase boundary (APB) energy. Such strength-microstructure relationships can improve calculations of deformation and failure in simulations that treat larger length and time scales, such as crystal plasticity (CP) simulations. 3D DDD simulations have also been advanced to track the formation and destruction of intrinsic and complex faults[27], processes that are strongly dependent on superalloy composition. Another noteworthy extension is the inclusion of misfit stresses resulting from significant lattice mismatch between two phases.[28,29]

**Using alloying to improved formability in magnesium**

The class of hexagonal close packed (HCP) metals, such as Mg, Zr, and Ti and their alloys, are being considered for a broad range of high-performance structural applications.[30] They bear many intrinsic properties that are attractive, such as low specific density, fatigue resistance, biocompatibility, corrosion resistance, and radiation resistance. Successful incorporation of HCP alloys into engineering designs is, however, hindered because their structural behavior is challenging to predict.

Mg alloys are one prominent and current example. For several decades, they have drawn attention as an ideal candidate for lightweight transportation due to low density and high specific strength.[31,32] However, their poor formability at room temperature limits its widespread use.[33] For Mg, there are several modes of slip, with the common slip modes being: basal <a>, prismatic <a>, and pyramidal <c+a> slip. Each one is distinct in its crystallographic slip plane and direction and critical resolved shear stress (CRSS) to activate it.[34,35] Poor formability is a consequence of the significant differences in slip mode CRSS.

Alloying can increase or diminish CRSS differences. In recent work, a multiscale polycrystal plasticity model (MPPM) was employed in an attempt to comprehensively explore solute effects on the CRSS values and in turn on formability.[36] At the highest length scale, the model utilizes a micromechanics formulation to relate the deformation of an aggregate of crystals to the deformation of an individual crystal. A grain-scale model for deformation twinning is used to account for division of a crystal into twinned and un-twinned crystalline domains. Each domain is in turn permitted to deform by crystallographic slip. Finally, the CRSS values for slip for a particular alloy are used as input and these are allowed to evolve with local amounts of slip strain.

CRSS differences among slip modes can be quantified via a plastic anisotropy (PA) measure. Presuming that basal slip is the easiest slip mode, which is typically the case for Mg alloys, it is given as:

$$PA = \frac{\tau_0^{Prismatic} - \tau_0^{Basal}}{\tau_0^{Twin} - \tau_0^{Basal}} \tag{2}$$

where $\tau_0$ is CRSS and the superscript indicates the slip or twin mode.

For a number of high performance Mg alloys ranging broadly in PA measure, MPPM was employed to simulate the plastic response and extract quantitative measures associated with formability. **Figure 5a** presents the calculated ratios of the tensile yield stress to the compressive yield stress. Overall the values are consistent with experimental measurements.[37-43] The important finding in **Figure 5a** is the strong relationship between the calculated tension-compression yield stress ratio and the plastic anisotropy PA measure.

As another measure of formability, the MPPM was used to calculate the polycrystal yield surfaces (PCYS) of many alloys. **Figure 5(b-c)** presents the calculated $\pi$-plane projection of the PCYS for pure Mg and the two different Mg alloys. A remarkable feature is the broad range of yield stresses that can be achieved via alloying. The Mg-4Li alloy has the smallest PCYS (**Figure 5(b)**) and the WE43 alloy the largest (**Figure 5(c)**). Asymmetries in the PCYSs reflect that plastic anisotropy, and, as anticipated, the level of asymmetry, scale with the PA measure. In revealing a strong correlation between the PA measure for slip and key measures for formability, MPPM introduced a new measure that can be used to screen for alloys that would potentially be formable.

**Hierarchical microstructure-sensitive structural properties**

The structural properties of today's most advanced alloys are highly dependent on a hierarchy of material microstructure, with length scales ranging from nanoscale (e.g., small precipitates, dislocations, interfaces) to microscale (e.g., large precipitates, dislocation cell substructure) and above (e.g., arrangement of phases and grains).[44] As we have seen thus far, Ni-based superalloys are one outstanding example. Within the mesoscale regime alone, their microstructures involve three length scales: (i) sub-grain scale, including the size/shape of $\gamma'$ precipitates and their spacing within the $\gamma$ matrix; (ii) grain scale, such as the size and crystallographic orientations of grains; (iii) polycrystal scale concerning how grains are aggregated.[45] Other families of high-performance alloys bearing such complex, multi-phase microstructures are Co-based superalloys, Al-Cu alloys, Mg-Ca-Zn alloys, and twinning-induced plasticity and transformation-induced plasticity steels.[7,20,46,47]

Many MMMs for the plastic deformation response of polycrystalline alloys employ CP theory. CP theory relates the distortions of a strained crystal to slip on crystallographic slip systems. Ideal for microstructure-sensitive calculations in alloys are combinations of CP with 3D full-field, spatially resolved mechanics techniques, such as CP finite element (CPFE) or CP fast Fourier transform (CP-FFT) solvers. To date, these MMMs have been advanced so that they span a wide range of scales, from non-uniform, time-varying applied fields (such as in mechanical shaping processes) to the sample, the multiphase microstructure at the mesoscale, the granular features such as texture and grain size within each phase at the microscale, and the crystal structure and associated operation of slip and twinning at the nanoscale. They explicitly couple the effects of microstructure morphology and crystallinity in the calculation of stress and strain evolution inside the grains and at the boundaries and interfaces.

Application of CPFE based modeling for Ni-based superalloys, including multiscale microstructures, can be found in many very recent works.[48,49] For instance, Keshavarz and Ghosh[50] developed a two-scale CPFE model to incorporate three mesoscale microstructures (**Figure 7**). At the sub-grain scale, mechanistic processes, such as dislocation evolution with non-Schmid effects and temperature dependence, APB shearing of $\gamma'$ precipitates, and micro-twinning. At the grain and polycrystal scales, they introduced an activation energy (AE)-based CP model that homogenized sub-grain scale responses and a model for geometrically necessary dislocations developed. The MMM was able to reproduce the tension-compression asymmetry and temperature-dependent slip mode transition, an unusual characteristic of these alloys that had been a challenge to predict.

**Grain neighborhood effects in alloys that twin**

Challenges remain in designing structural HCP alloys since they can potentially deform not only by slip but also by deformation twinning. Deformation twinning is receiving a lot of attention, since it is not as well understood as plastic slip and has a stronger effect on the mechanical response when it happens.[51] When the alloy is strained, atomic-scale twins can form inside grains or grain boundaries and grow a few orders of magnitude to span the grain. The twin domain abruptly reorients and shears the lattice.

Alloying greatly affects the propensity of twinning and plays a significant role in the twin-architecture that develops with straining. Twins that transmit across grain boundaries can detrimentally lead to the formation of "twin chains" percolating across the sample and triggering fatigue cracks and premature failure.[52-54] Twins that instead remain in their parent grain and multiply in the form of 3D intersecting networks can favorably lead to simultaneous high strength and large compression strain to failure.[55] How alloying controls twin development is largely unknown.

In the last few years, multiscale modeling tools for discrete twin domains in this mesoscale regime have been developed.[56-59] These techniques enable calculations of the local stress fields and dislocation activity around twin lamellae, shedding light on the effects of size and local grain neighborhoods on the propensity for twin growth. In one study, this type of multiscale model based on CP-FFT was employed to study the effect of alloying on the transmission of twins across grain boundaries.[60] **Figure 6(a-b)** plots the ratio of a driving force for transmission into the neighboring crystal across the boundary to that for growth of the same twin in its own parent crystal with respect to the misorientation of the neighboring grain. Two Mg alloys are considered, and in both, the driving force ratio rapidly decreases from unity at zero misorientation (no grain boundary) to zero as the misorientation between the two crystals increases, a trend that would be expected from purely geometric arguments. The important finding is the strong influence of alloying on the cut-off misorientation angle $\Delta\theta_{cut}$, above which the driving force is zero and chances for twin transmission is nil. The alloy with the higher PA (**Equation 2**), AZ31, has a much higher $\Delta\theta_{cut}$ than Mg4Li, with the lower PA. Evidently, alloys with larger CRSS gaps among their slip modes promote twin transmission.

**Figure 6(c)** maps the calculated $\Delta\theta_{cut}$ for a wide range of Mg alloys. For more anisotropic (higher PA) alloys, $\Delta\theta_{cut}$ increases with PA, while for lower PA alloys, $\Delta\theta_{cut}$ is fixed at a lower value of ~50°. These multiscale CP-FFT predictions reveal that alloying can have a more direct effect on twin morphologies than first thought. The insight suggests that alloying can help to prevent twin chains from forming and acting as preferred paths for shear banding or cracking.

**Opportunities and Challenges**

As we have highlighted, significant advances have been made in linking composition, microstructure, and mechanisms with the development of local stress states and deformation response. Yet several multiscale modeling challenges remain in the quest to fully understand processing-microstructure-response relationships of an alloy.

During deformation, particularly under elevated temperatures and long periods of time, a number of phenomena can occur that involve stress-induced migration of internal boundaries, such as phase transformations, recrystallization, grain growth, crack growth, and deformation twins. Many of these aspects have been studied intensively using atomic scale simulations and PF methods, but incorporating their effects on slip and twinning in MMM codes has been challenging. Recent advances in this area have involved concurrently combining PF and CP theory into a single simulation tool, allowing for updates of crystalline and thermodynamic properties of the phases, lattice defects, and kinetics of migrating interfaces in time or strain.[61-63] These have been used to understand the influence of dislocation dynamics on rafting and how microstructural evolution can constrain or facilitate dislocation activity.

Many new generations of ultra-strong alloys have a highly heterogeneous grain structure. Some microstructural regions consist of micron-size grains, while other regions are nanostructured, comprised of nanotwins, bimetal nanolayers, metal-ceramic nanolaminates, and nanograins.[64,65] No model to date can treat the plasticity in a such a significantly varied grain structure. Current multiscale models treat plasticity in coarse grains by scale-independent, statistical dislocation densities over long time scales, and in nanograins by the scale-dependent motion of discrete dislocations over short time scales. New methodologies that can address discrete slip occurring over long times, in which microstructural evolution can be captured, are needed to design newer forms of heterogeneous alloys. Some recent advances to implement discrete slip events into a CPFE framework have applied to relatively simple single-phase, pure metal systems.[66]

Many high-performance alloys are multi-phase, wherein more than one phase can plastically deform. Atomic-scale simulation and *in-situ* microscopy of deforming groups of grains have revealed a number of defect/interface reactions (transference, recovery, nucleation) that could significantly affect the types of slip and twinning modes that would be selected in deformation.[67,68] Representing the role of such highly resolved atomic-scale reactions into the local orientation of crystals and ultimately the mechanical response is beyond current modeling capabilities. To benefit the advancement of alloys by microstructure design and control, MMM extensions toward incorporating the role of dislocation/interface interactions is recommended.

**Outlook**

Overcoming the "mesoscale gap" is rapidly becoming an MMM community level effort. Notable center-level efforts exist that have successfully connected methodologies that span the broad atoms-to-continuum time and length scale spectrum.[69,70] Data I/O, codes, and reports are increasingly being shared in repositories, publications, and public websites and hubs worldwide.[71,72] In tandem are growing efforts to train users, via on-line manuals, example cases, summer schools, and workshops.[73,74] Leaps in computing power are undeniably indispensable to fully develop and realize the predictive capabilities of many MMMs. Promising ways toward enabling larger and longer time simulations include adopting novel computing architectures, innovating efficient computational schemes, and using government and national laboratory High Performance Computing (HPC) facilities[75-78], and DOE's Exascale Project.[79] The Minerals, Metals & Materials Society (TMS) recently published two reports, Modeling Across Scales[80] and Core Knowledge and Skills[81], in order to inform the public of methodology and knowledge gaps. By coming together, MMMs for alloys can truly conquer the length scales within our lifetime.

**Acknowledgments**

IJB and SX gratefully acknowledge support in part from the Office of Naval Research under contract ONR BRC Grant N00014-18-1-2392. The work of SX was supported in part by the Elings Prize Fellowship in Science offered by the California NanoSystems Institute (CNSI) on the UC Santa Barbara campus. JL acknowledges support by the European Research Council (ERC) under the European Union's Horizon 2020 research and innovation program (Advanced Grant VIRMETAL, grant agreement No. 669141). JAE acknowledges support from the NSF CAREER Award #CMMI-1454072 and ONR Award #N00014-18-1-2858. JRM and BS gratefully acknowledge financial support of the work through Project M5, Priority Program 1713 (Chemo-mechanics), German Science Foundation (DFG), and through EPSRC Program Grant EP/R001715/1 (LightForm).

## References

1. M.F. Ashby, D. Cebon, *J. Phys. IV France* **3**, C7-1 (1993).

2. E.O. Hall, *Proc. Phys. Soc. Lond.* **64**, 747 (1951).

3. N.J. Petch, *J. Iron Steel Inst. London.* **173**, 25 (1953).

4. W. Hume-Rothery, H. M. Powell, *Z. Krist.* **91**, 23 (1935).

5. J. Nie, Y. Wang, Microstructural design for advanced light metals, *MRS Bull.* (2019)

6. A. Rodríguez-Veiga, B. Bellón, I. Papadimitriou, G. Esteban-Manzanares, I. Sabirov, J. LLorca, *J. Alloys Comp.* **757**, 504 (2018).

7. H. Liu, I. Papadimitriou, F. X. Lin, J. LLorca, *Acta Mater.* (submitted).

8. H. Liu, B. Bellón, J. LLorca, *Acta Mater.* **132**, 611 (2017).

9. G. Esteban-Manzanarez, E. Martínez, J. Segurado, L. Capolungo, J. LLorca, *Acta Mater.* **162**, 189 (2019).

10. U. F. Kocks, Prog. Mater. Sci. **19**, 1 (1975).

11. R. Santos-Güemes, G. Esteban-Manzanares, I. Papadimitriou, J. Segurado, L. Capolungo, J. LLorca, *J. Mech. Phys. Solids* **118**, 228 (2018).

12. Y. Koizumi, T. Nukaya, S. Takeshi, S. Suzuki, S. Kurosu, Y. Li, H. Matsumoto, K. Sato, Y. Tanaka, A. Chiba, *Acta Mater.* **60**, 2901 (2012).

13. G.B. Viswanathan, R. Shi, A. Genc, V.A. Vorontsov, L. Kovarik, C.M.F. Rae, M.J. Mills, *Scr. Mater.* **94**, 5 (2015).

14. Y. Rao, T.M. Smith, M.J. Mills, M. Ghazisaeidi, *Acta Mater.* **148**, 173 (2018).

15. P. Kontis, Z. Li, D.M. Collins, J. Cormier, D. Raabe, B. Gault, *Scr. Mater.* **145**, 76 (2018).

16. Y. Wang, J. Li, *Acta Mater.* **58**, 1212 (2010).

17. I.J. Beyerlein, A. Hunter, *Philos. Trans. A* **374**, 20150166 (2016).

18. J.R. Mianroodi, A. Hunter, I.J. Beyerlein, B. Svendsen, *J. Mech. Phys. Solids* **95**, 719 (2016).

19. R. Shi, D.P. McAllister, N. Zhou, A.J. Detor, R. DiDomizio, M.J. Mills, Y. Wang, *Acta Mater.* **164**, 220 (2019).

20. J.R. Mianroodi, P. Shanthraj, P. Kontis, B. Gault, D. Raabe, B. Svendsen, under review (2018).

21. B. Svendsen, P. Shanthraj, D. Raabe, *J. Mech. Phys. Solids* **112**, 619 (2018).

22. L.P. Kubin, G. Canova, M. Condat, B. Devincre, V. Pontikis, Y. Bréechet, *Solid State Phenom.* **23-24**, 455 (1992).

23. N.M. Ghoniem, S.-H. Tong, L.Z. Sun, *Phys. Rev. B* **61**, 913 (2000).

24. H.M. Zbib, M. Rhee, J.P. Hirth, *Int. J. Plast.* **18**, 1133 (2002).

25. D. Weygand, L.H. Friedman, E. Van der Giessen, A. Needleman, *Model. Simul. Mater. Sci. Eng.* **10**, 437 (2002).

26. J.A. El-Awady, H. Fan, A.M. Hussein, in: *Multiscale Materials Modeling for Nanomechanics*, C. Weinberger, G. Tucker, Ed. (Springer, Cham, 2016) pp. 337-371.

27. A.M. Hussein, S.I. Rao, M.D. Uchic, T.A. Parthasarathy, J.A. El-Awady, *J. Mech. Phys. Solids* **99**, 146 (2017).

28. H. Yang, Z. Li, M. Huang, *Comp. Mater. Sci.* **75**, 52 (2013).

29. M. Huang, L. Zhao, J. Tong, *Int. J. Plast.* **28**, 141 (2012).

30. S. Gao, M. Fivel, A. Ma, A. Hartmaier, *J. Mech. Phys. Solids* **76**, 276 (2015). C.N. Tomé, I.J. Beyerlein, R.J. McCabe, J. Wang, in: *Engineering (ICME) for Metals: Reinvigorating Engineering Design with Science*, M.F. Horstemeyer, Ed. (Wiley Press, 2018) pp. 283-336.

31. N.J. Kim, *Mater Sci. Tech.* **30**, 1925 (2014).

32. M.K. Kulekci, *Int. J. Adv. Manuf. Technol.* **39**, 851 (2008).

33. B. Suh, M.S. Shim, K.S. Shin, N.J. Kim, *Scr. Mater* **84-85**, 1 (2014).

34. P.G. Partridge, *Metall. Rev.* **12**, 169 (1967).

35. M.H. Yoo, *Met. Trans.* **124**, 409 (1981).

36. M. Arul Kumar, I.J. Beyerlein, C.N. Tomé, *J. Alloys Compd.* **695**, 1488 (2017).

37. M. Lentz, M. Klaus, R.S. Coelho, N. Schaefer, F. Schmack, W. Reimers, B. Clasuen, *Metall. Mater. Trans.* **45A**, 5721 (2014).

38. H. Qiao, S.R. Agnew, P.D. Wu, *Int. J Plast.* **65**, 61 (2015).

39. S. Xu, T. Liu, H. Chen, Z. Miao, Z. Zhang, W. Zeng, *Mater Sci. Eng. A* **565**, 96 (2013).

40. W. Muhammad, M. Mohammadi, J. Kang, R.K. Mishra, K. Inal, *Int. J. Plast.* **70**, 30 (2015).

41. P. Zhou, E. Beeh, H.E. Friedrich, *J. Mater. Eng. Perform.* **25**, 853 (2013).

42. Z. Zachariah, Sankara Sarma V. Tatiparti, S.K. Mishra, N. Ramakrishnan, U. Ramamurty, *Mater Sci. Eng. A* **572**, 8 (2013).

43. S. Yi, J. Bolen, F. Heineman, D. Letzig, *Acta Mater.* **58**, 592 (2010).

44. D.L. McDowell, in: *Computational Materials System Design*, D. Shin, J. Saal, Ed. (Springer, Cham, 2018) pp. 1-25.

45. S. Keshavarz, S. Ghosh, *Int. J. Solids Struc.* **55**, 17 (2015).

46. A.A. Luo, *Int. Mater. Rev.* **49**, 13 (2004).

47. B.C. De Cooman, Y. Estrin, S.K. Kim, *Acta Mater.* **142**, 283 (2018).

48. A. Bagri, G. Weber, J.C. Stinville, W.C. Lenthe, T.M. Pollock, C. Woodward, S. Ghosh, *Metall. Mater. Trans. A* **49**, 5727 (2018).

49. M. Pinz, G. Weber, W.C. Lenthe, M.D. Uchic, T.M. Pollock, S. Ghosh, *Acta Mater.* **157**, 245 (2018).

50. S. Keshavarz, S. Ghosh, *Int. J. Solids Struc.* **55**, 17 (2015).

51. I.J. Beyerlein, M. Arul Kumar, in: *Handbook of Materials Modeling*, W. Andreoni, S. Yip, Ed. (Springer Nature, 2018) pp. 1-36.

52. B.A. Simkin, M.A. Crimp, T.R. Bieler, *Intermetallics* **15**, 55 (2007).

53. F. Yang, S. M. Yin, S. X. Li, Z. F. Zhang, *Mater. Sci. Eng. A* **491**, 131 (2008).

54. S.M. Yin, F. Yang, X.M. Yang, S.D. Wu, S.X. Li, G.Y. Li, *Mater. Sci. Eng. A* **494**, 397 (2008).

55. M. Lentz, M. Risse, N. Schaefer, W. Reimers, I.J. Beyerlein, *Nature Comm.* **7**, 11068 (2016)

56. J. Cheng, S. Ghosh, *J. Mech. Phys. Solids* **99**, 512 (2017).

57. H. Abdolvand, A.J. Wilkinson, *Acta Mater.* **105**, 219 (2016).

58. M. Ardeljan, I.J. Beyerlein, M. Knezevic, *Int. J. Plast.* **99**, 81 (2017).

59. M. Arul Kumar, I.J. Beyerlein, C.N. Tomé, *Acta Mater.* **116**, 143 (2016).

60. M.A. Kumar, I.J. Beyerlein, R.A. Lebensohn, C.N. Tome, *Mater. Sci. Eng. A* **706**, 295 (2017).

61. M. Cottura, B. Appolaire, A. Finel, Y. Le Bouar, *J. Mech. Phys. Solids* **94**, 473 (2016).

62. R. Wu, S. Sandfeld, *J. Alloys Compd.* **703**, 389 (2017).

63. R. Wu, M. Zaiser, S. Sandfeld, *Int. J. Plast.* **95**, 142 (2017).

64. Z. Li, K.G. Pradeep, Y. Deng, D. Raabe, C.C. Tasan, *Nature* **534**, 227 (2016)

65. T. Xiong, Y. Zhou, J. Pang, I.J. Beyerlein, X. Ma, S. Zheng, *Mater. Sci. Eng. A* **720**, 231 (2018).

66. R. Yuan, I.J. Beyerlein, C. Zhou, *Acta Mater.* **110**, 8 (2016).

67. I.J. Beyerlein, X. Zhang, A. Misra, *Annu. Rev. Mater. Res.* **44**, 329 (2014).

68. I.J. Beyerlein, M.J. Demkowicz, A. Misra, B.P. Uberuaga, *Prog. Mater. Sci.* **74**, 125 (2015).

69. http://www.prisms-center.org/

70. https://www.questek.com/

71. https://nanohub.org/

72. https://matin.gatech.edu/

73. https://magics.usc.edu/

74. https://cms3.tamu.edu/

75. http://www.nersc.gov/

76. https://www.xsede.org/

77. https://hpcinnovationcenter.llnl.gov/

78. https://usrc.lanl.gov/

79. https://www.exascaleproject.org/

80. The Minerals, Metals & Materials Society (TMS), *Modeling Across Scales: A Roadmapping Study for Connecting Materials Models and Simulations Across Length and Time Scales* (Warrendale, PA, 2015).

81. The Minerals, Metals & Materials Society (TMS), *Advanced Computation and Data in Materials and Manufacturing: Core Knowledge Gaps and Opportunities* (Pittsburgh, PA, 2018).

**Figure and Table Captions**

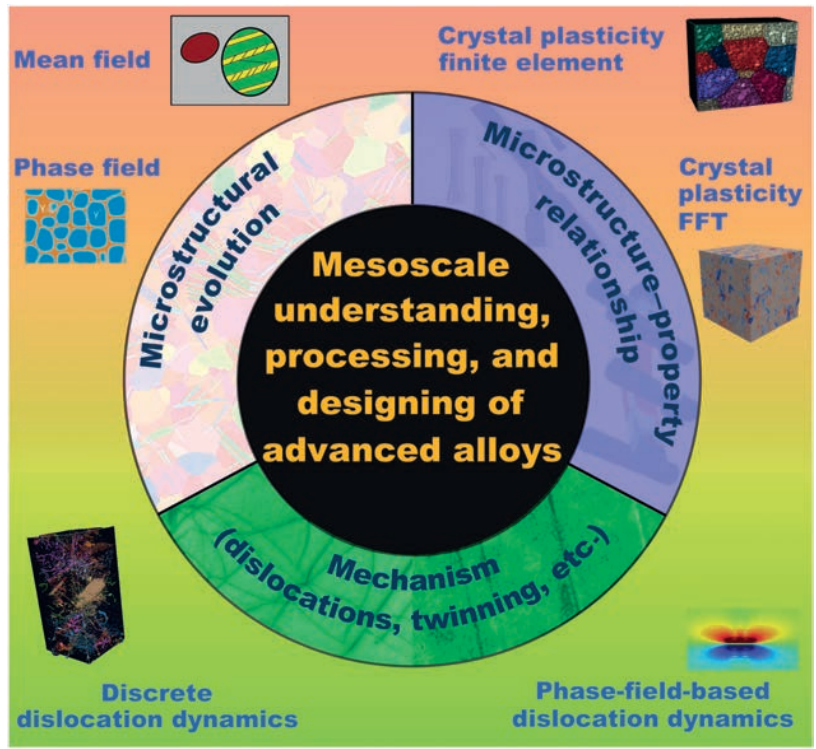

**Figure 1.** Current view on the state of mesoscale multiscale materials modeling.

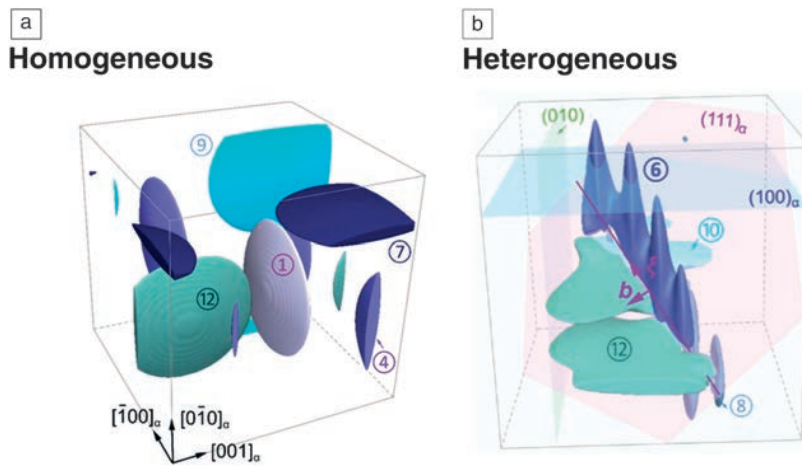

**Figure 2.** Final precipitate size and morphologies predicted from multiscale simulations elucidating the differences that can be expected in high temperature precipitate homogeneous and heterogeneous nucleation and growth in Al-Cu alloys. Reproduced with permission from Ref. [8].

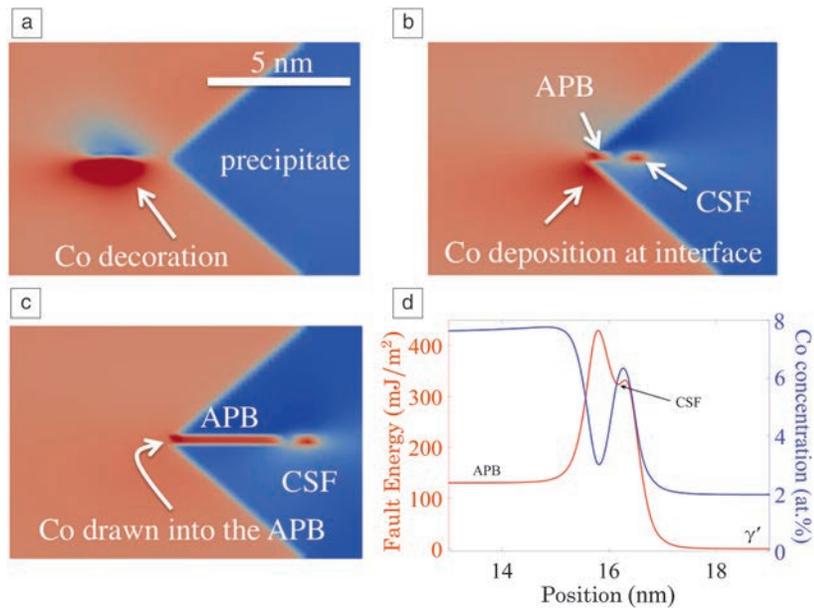

**Figure 3.** (a, b, c) Shear of $\gamma'$ precipitate by Co decorated $\langle 110 \rangle/2$ edge dislocation in $\gamma$. Dark blue and dark red represent 3.2 and 6.0 at.% Co, respectively. CSF stands for "complex stacking fault", APB for "antiphase boundary". All faults are separated by partial dislocations. (d) Fault energy results for a dissociated $\langle 110 \rangle/2$ in Ni$_3$Al-Co (red curve). Segregated Co concentration profile shown in blue. Results in (a, b, c) are for a moving dislocation, and those in (d) for a static dislocation. See text for details. Reproduced with permission from Ref. [20].

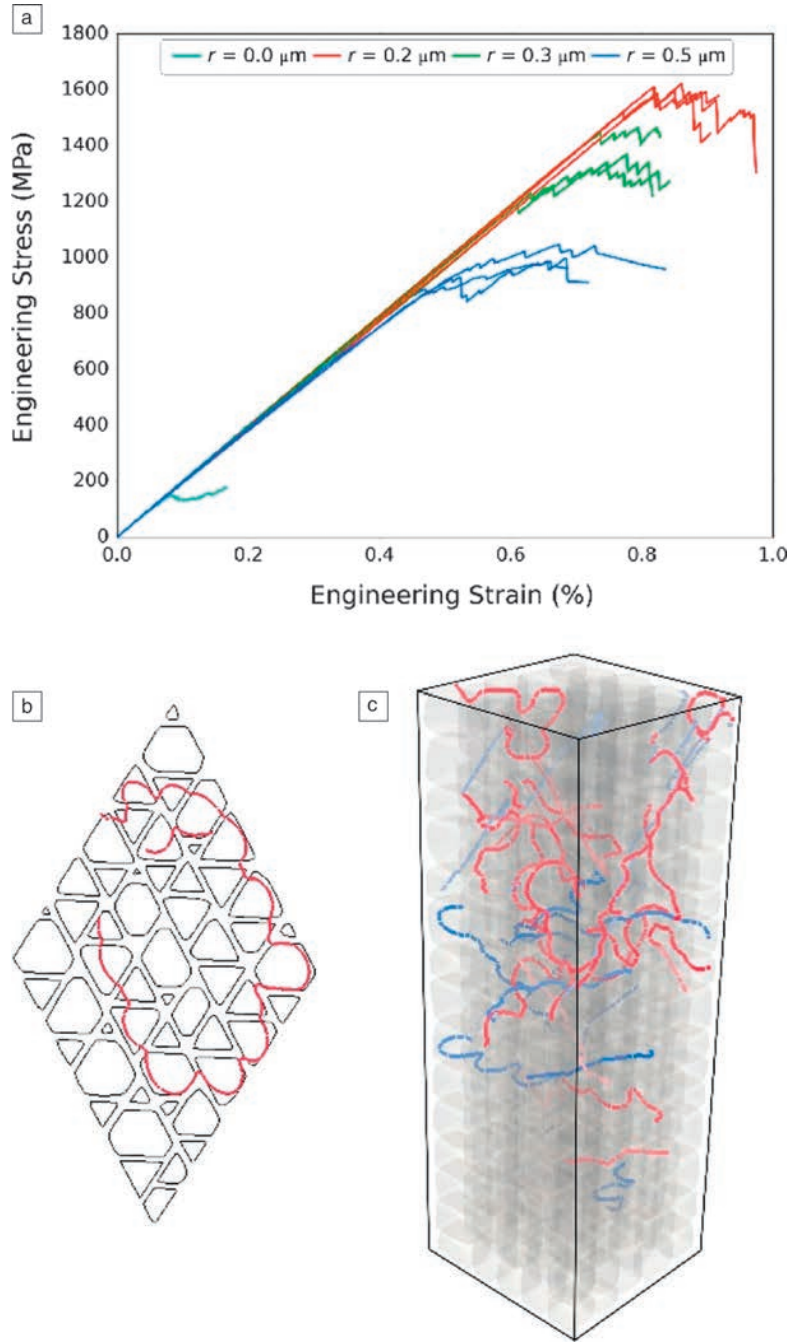

**Figure 4.** The effect of the precipitate size, *r*, on the (a) engineering stress-strain response; and (b) dislocation density versus engineering strain, for free-standing single crystal Ni-base superalloy microcrystals. The simulation cell size was fixed at 1×1×3 μm³, the precipitates were cuboidal with a volume fraction of 0.7, and the APB energy was 0.2 J/m². The precipitate distribution and the dislocation network at the onset of the plastic flow are shown in (b) and (c) for *r* = 0.2 μm and 0.5 μm, respectively. Thin slices extracted parallel to the (111) slip plane from the center of each simulation cell are also shown. The dislocation lines in (b) and (c) are colored according to their slip system. Reproduced with permission from Ref. [27].

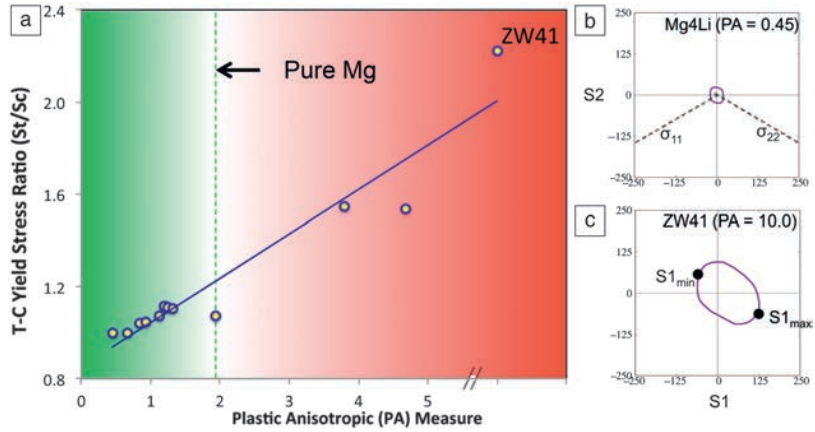

**Figure 5.** (a) Effect of alloying addition on tension-compression yield stress ratio for magnesium alloys. Magnesium alloys with different alloying elements are characterized by plastic anisotropic measure. (b-c) Polycrystalline yield surface (π-plane projection) for different magnesium alloys at 3% compressive strain along the rolling direction. Reproduced with permission from Ref. [36].

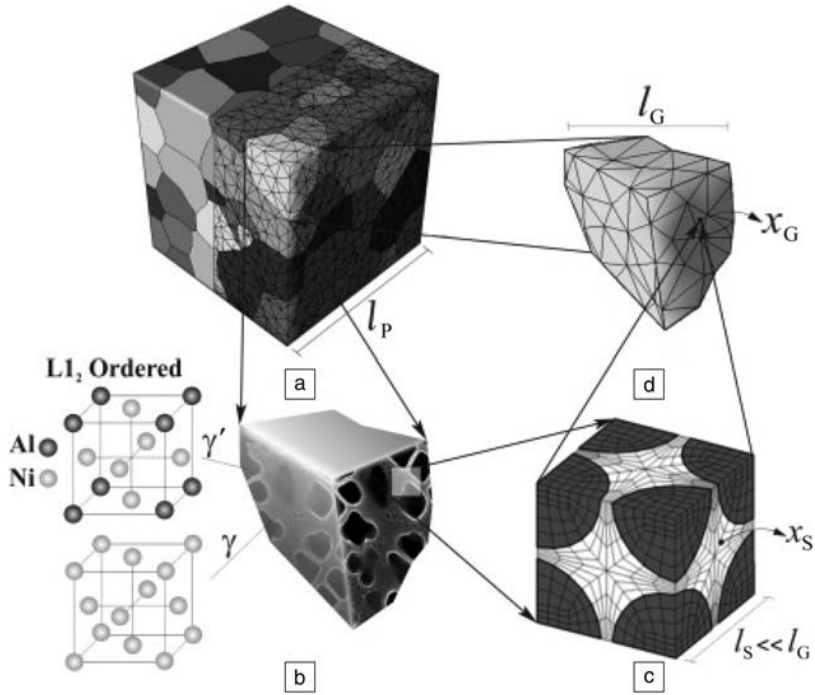

**Figure 6.** Variation of the driving force for twin transmission across a grain boundary with the misorientation of the grain boundary for two Mg alloys (a) AZ31 and (b) Mg4Li. A value of unity implies that the driving force for twinning in neighboring grain at the tip of the twin is equal to that for twin propagation of the same twin in its parent crystal. A high (or too high) misorientation regime is well marked by a cut-off misorientation angle $\Delta\theta_{cut}$, above which the chances for twin

transmission are zero. (c) Map of the variation of cut-off angles for a wide range of alloys. Alloys are indicated by their PA measure in **Equation 2**. An arbitrarily defined practical cut off angle, associated with a driving force ratio of 0.5 is also marked. This angle corresponds to a neighboring grain that bears a driving force that is 50% lower than that to propagate a twin in its own parent crystal. These more practical cut-off angles follow the same variation with PA as seen in (c). However, this 50%-chance misorientation angle is more likely to be consistent with experimental observation. It is reasonable to expect that twin transmission would rarely be seen, particularly when the datasets are small, for misorientations corresponding this practical cut-off angle, which is ~36° for the low PA alloys and ~50° for the highest one. Reproduced with permission from Ref. [60].

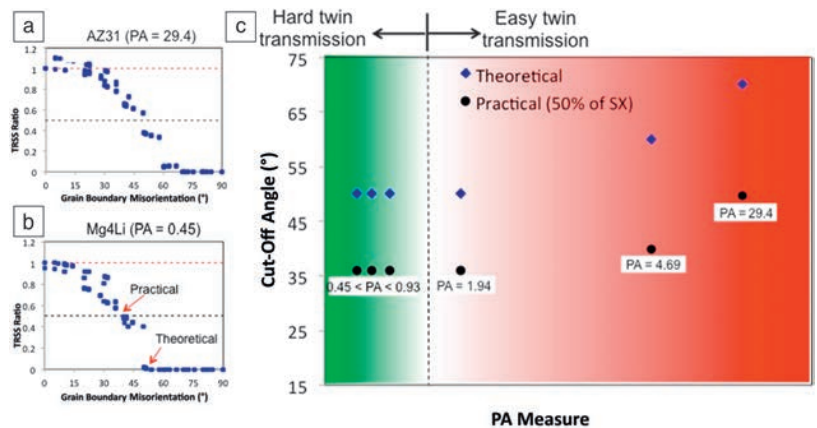

**Figure 7.** Schematic representation of multiple scales in the development of a CPFE model for Ni-based superalloys: (a) polycrystalline microstructure showing the finite element mesh, (b) subgrain microstructure in a single grain, (c) discretized subgrain microstructural RVE, and (d) homogenized CPFE model for a grain. Reproduced with permission from Ref. [50].

## Author biographies

Type a brief (~80 words) biography for each author here. Include current employment title and institution; research interests; education (degrees, institutions, and fields of study); highlights of employment history; and relevant accomplishments, awards, publications, organizational involvement, etc., as appropriate. At the beginning of the biography, include full contact information (mailing address, telephone, fax, and e-mail). Remember to include a high-resolution head-and-shoulders color photograph for each author when submitting the final manuscript. Guidelines for submitting your author photo are located here.